\documentclass[12pt,epsf]{article}
\usepackage{epsfig}
\usepackage{amsmath}
\usepackage{amsfonts}
\usepackage{amssymb}
\def\R{{\rm I\!R}}
\newtheorem {Definition}{Definition}
\newtheorem {Proposition}{Proposition}

\begin{document}
\title{Kontsevich and Takhtajan construction of star product on the Poisson 
Lie group $GL(2)$} 
\author{N. Bel Baraka${^{1,2}}$\\
{\small ${^1}$ Laboratoire Gevrey de Math\'ematique Physique,CNRS UMR 5029,}\\ 
{\small Universit\'e de Bourgogne,
 BP 47870, F-21078 Dijon~~ France}\\
{\small ${^2}$ Laboratoire de Physique Th\'eorique,
~Universit\'e MohammedV},~~\\
{\small BP 1014,~~Rabat~~Maroc}}
\maketitle
{\bf Abstract}:  
Comparing the star product defined by Takhtajan on the 
Poisson-Lie group $GL(2)$ and any star product calculated from the 
Kontsevich's graphs (any ''K-star product'') on the same group, we show,
by direct computation, that the Takhtajan star product on $GL(2)$
can't be written as a K-star product.

\vfill

\section{Introduction} 

In recent years great progress was made in developing new approach 
and deriving exact result in deformation of different groups and algebras. Each
of these deformation theories is not independent of the others.\\
In fact, since the Kontsevich's well known preprint \cite{r2}, in which 
he gives an universal construction of a star product on $\R^{d}$
endowed with an arbitrary Poisson structure, several authors tempted to 
bring this approach closer to others already existing, let us cite, for instance,
 D. Arnal, N. Ben Amar and M. Masmoudi \cite{r6} and G. Dito \cite{r7} who give 
by two different manners an equivalence between the Kontsevich and Gutt \cite{r5}
star product on the dual of Lie algebra, and V.Kathotia \cite{r8}
and B. Shoikhet \cite{r9} who related the Kontsevich formula to 
Campbell-Baker-Hausdorff's one on the dual of Lie algebra.\\
The starting point of the present idea is the Drinfeld universal 
approach to construct quantum groups \cite{r3}, this mathematical
structure arises in particular from quantization of some Poisson bracket
on ''usual'' Lie groups obtained from a classical $r$-matrix satisfying the 
Yang Baxter Equation. Here, we tempt to illimunate the relation between
the star product construct by Takhtajan, (basing on Drinfeld's work),
on the particular Lie group $GL(2)$ endowed with a certain $r$-matrix
which satisfies the modified Yang Baxter Equation, and the star product 
constructed on this Poisson Lie group from the Kontsevich's graphs (''K-star 
product'') either on $GL(2)$ view as an open subset of $\R^{4}$ or on the 
domain of an exponential chart near the origin. 
By a direct computation, we show that the Takhtajan 
star product can not be written as a K-star product.\\
This paper is organized as follows, the second section is devoted to
a review of basic definitions of the quantization of Poisson Lie group, 
the third section introduce the Kontsevich construction, in section 4 
we give a generalization of this construction, then we give the 
quantization of the particular Poisson Lie group $GL(2)$ in the fifth
section, finally, we get our main result by comparing the two star
product on an ''ordinary'' and an ''exponential'' chart in the three last
sections.

\section{Usual quantization of Poisson Lie group:}  

 Let us first recall the aim of the construction of quantum groups by V. Drinfeld and L. Takhtajan \cite{r1,r3}.
Let $G$ be a Lie group with Lie algebra $g$, we denote by $(X_{i})$ a basis 
of $g$ and $U(g)$ 
the universal enveloping algebra of $g$.
If $r \in \wedge^{2}g$, we consider the elements $r^{12}, r^{13}, r^{23}$
of $ U(g)\otimes U(g)\otimes U(g)$ definded by:
$$ r^{12} = r^{ij} X_{i} \otimes X_{j} \otimes 1$$
$$ r^{13} = r^{ij} X_{i} \otimes 1 \otimes X_{j} $$
$$ r^{23} = r^{ij} 1 \otimes X_{i} \otimes X_{j}$$
where $ r = r^{ij} X_{i} \otimes X_{j}$. We say that $r$ satisfies the modified 
Classical Yang-Baxter Equation (CYBE) if:

\begin{equation} 
\lbrack r^{12}, r^{13}\rbrack +  \lbrack r^{12}, r^{23}\rbrack + \lbrack r^{13}, 
r^{23}\rbrack  =  I_{123} ,~~~~I_{123} \in \wedge^{3} g
\end{equation}\\
and
\begin{equation}
\lbrack I_{123}, 1\otimes1\otimes X + 1\otimes X\otimes1 + 
X\otimes1\otimes1\rbrack = 0  ~~~~ \forall X \in g
\end{equation}\\
(here the bracket is the commutator in the associative algebra
$U(g)\otimes U(g)\otimes U(g)$). 
 Such an element is called a $r$-matrix.\\
To each $r$, we associate a Poisson structure on $G$ by putting:

\begin{equation} 
\{\varphi, \psi \} = r^{ij}(X_{i}^{\ell}(\varphi)X_{j}^{\ell}(\psi)
- X_{i}^{r}(\varphi)X_{j}^{r}(\psi))~~~~\varphi, \psi \in C^{\infty}(G)
\end{equation}\\
where $X_{i}^{\ell}(resp. ~X_{j}^{r})$ are the left-invariant (resp. 
right-invariant) vector fields on $G$ corresponding to $X_{i}(resp. ~X_{j})$.\\
\begin{Definition} (Poisson Lie group)\\
A Poisson-Lie group is a Lie group $G$ endowed with a Poisson structure $\{, \}$ 
associated to a $r$-matrix satisfying the modified CYBE.
\end{Definition}
\vspace{.3in}
The quantization of a Poisson-Lie group $(G, \{, \})$ is a deformation of 
the commutative algebra $C^{\infty}(G)$ which turns it to a new noncommutaive
algebra $C^{\infty}(G)[\![ t ]\!]$, where $t$ is a deformation parameter.
The algebra $C^{\infty}(G)[\![ t ]\!]$ as a vector space coincides with 
$C^{\infty}(G)$, but has a new product $\ast$ called a star product.\\
\begin{Definition} (Star product)\\
A star product on a Poisson manifold is a map:
$$ \ast : C^{\infty}(G) \otimes C^{\infty}(G) \longrightarrow C^{\infty}(G)[\![ t ]\!]$$
$$ \varphi \ast \psi = \varphi.\psi + \sum_{i=1}C_{i}(\varphi, \psi)t^{i}$$\\
such that, for all $\varphi, \psi, \chi \in C^{\infty}(G)$:\\
$1)~C_{i}$ is a bidifferential operator on $C^{\infty}(G)$\\

\noindent $2)~\varphi \ast 1 = 1 \ast \varphi = \varphi $\\

\noindent $3)~\{\varphi, \psi \} = \lim_{t \rightarrow 0} \frac{1}{t} (\varphi \ast \psi - \psi \ast \varphi)$\\

\noindent $4)~(\varphi \ast \psi) \ast \chi = \varphi \ast (\psi \ast \chi) $.\\

\end{Definition}
Since $G$ is a group, there is a naturel comultiplication $\Delta$ on 
$C^{\infty}(G)$:
$$ \Delta(\varphi)(x, y) = \varphi(xy)~~~~(\varphi \in C^{\infty}(G), x, y \in G).$$\\
A star product preserving $\Delta$, i.e. such that:
\begin{equation} 
\Delta (\varphi \ast \psi ) = \Delta (\varphi ) \ast \Delta (\psi)
\end{equation}\\
where  $\ast$ is naturally extended to $C^{\infty}(G)\otimes C^{\infty}(G)$,
was built by V. Drinfeld and L. Takhtajan \cite{r1} in a purely algebraic way.
They first look for a formal element $F \in U(g) \otimes U(g)[\![ t ]\!]$
such that the product:
\begin{equation}\label{st}
\varphi \ast \psi = (F^{-1})^{r}(F)^{\ell}(\varphi \otimes \psi)
\end{equation}\\
is a star product.
And the associativity axiom looks:
\begin{equation}\label{F} 
 F(X+Y,Z)F(X,Y) = \alpha(X,Y,Z) F(X,Y+Z)F(Y,Z)
\end{equation}
where $\alpha(X,Y,Z)\in U(g)\otimes U(g)\otimes U(g)[\![ t ]\!]$ is G-invariant:
\begin{equation} 
[ \alpha, 1\otimes 1 \otimes X + 1\otimes X \otimes1 + X\otimes 1 \otimes1 ] = 0~~~~ \forall X \in g.
\end{equation}\\
 In order to have this, we need that:
$$ F = 1 - \frac{t}{2}r + \sum_{n\geq2} F_{n}t^{n}$$\\
and $$ F(X,0) = F(0,Y) = 1 $$
this implies that $\alpha$ has the following form:
$$ \alpha = 1 + t^{2}\alpha_{2} + \cdots $$\\
with $$ Alt(\alpha_{2}) = -4I_{123}.$$\\
Here $Alt$ stands for the alternation, i.e.:\\
\begin{equation*}
\begin{split}
Alt(\alpha_{2})(X,Y,Z) &= \alpha_{2}(X,Y,Z)-\alpha_{2}(Y,X,Z)+\alpha_{2}(Y,Z,X)\\
&-\alpha_{2}(Z,Y,X)+\alpha_{2}(Z,X,Y)-\alpha_{2}(X,Z,Y) 
\end{split}
\end{equation*}
and$$\alpha(X,Y,Z)\alpha(X,Y+Z,U)\alpha(Y,Z,U) = 
\alpha(X+Y,Z,U)\alpha(X,Y,Z+U).$$\\
An explicit solution for $GL(2)$ will be given later.

\section{Kontsevich's star product on $\R^{d}$:}
In order to construct a star product on any Poisson manifold, 
M. Kontsevich built first such a star product for any Poisson structure 
$\Lambda$ on a flat space $\R^{d}$ with a given system of coordinates.\\
He considers a set $G_{n,m}$ of graphs $\Gamma$ with two kinds of vertices:
$n$ aerial vertices $p_{1},p_{2},......p_{n}$ and $m$ terrestrial vertices 
$q_{1}<q_{2}<.....<q_{m}$.
From each aerial vertex $p_{i}$, two edges (arrows) $\vec a_{i}$ are starting, 
they end at any different vertices $(a)$ distincts from $p_{i}$ (i.e. there are not parallel multiple
edges either ''small'' loops); on the edges, we fix the lexicographic ordering, we associate to the
graph $\Gamma$ a $m$-differential operator:
\begin{equation}
\begin{split}
B_{\Gamma}(\Lambda\otimes\Lambda\otimes\Lambda)(\varphi_{1},\varphi_{2},.....\varphi_{m}) &= 
\sum D_{p_{1}}\Lambda_{1}^{i_{1}i_{2}....i_{k_{1}}}.......D_{p_{n}}\\
&\Lambda_{n}^{i_{k_{1}+...k_{n-1}+1}...i_{k_{1}+...k_{n}}}
D_{q_{1}}\varphi_{1}....D_{q_{m}}\varphi_{m}
\end{split}
\end{equation}\\
where $D_{a}$ is the operator:
$$ D_{a} = \prod_{l,edge(l)=\vec{.a}} \partial_{i_{l}}.$$
Kontsevich looks for a star product of the form:
\begin{equation}
\varphi \ast \psi = \varphi.\psi + \sum_{n\geq1} t^{n}\sum_{\Gamma\in G_{n,2}} 
a_{\Gamma} B_{\Gamma}(\Lambda,\Lambda,.....\Lambda)(\varphi\otimes\psi)
\end{equation}\\
where $a_{\Gamma}$ is a constant.
An explicit universal choice of the $a_{\Gamma}$ is given in \cite{r2}, $a_{\Gamma}$ is the integral
of a certain form $\omega_{\Gamma}$ defined from $\Gamma$ on a configuration space $C^{+}_{n,2}$.
with this choice for any Poisson structure $\Lambda$, the star product of 
Kontsevich satisfies the conditions 1), 2), 3) and 4) of 
the preceding definition.

\section{Generalization of the Kontsevich construction:}
We first generalize the construction of Kontsevich on $\R^{d}$.\\
Let us consider now graphs with $n$ aerial vertices $p_{1},p_{2},......p_{m}$ 
and $m$ terrestrial vertices $q_{1},q_{2},.....,q_{m}$ and 2 edges starting from
 each aerial vertex $p_{i}$ and ending at any vertex (even possibly in $p_{i}$ 
 itself) without any double edge.\\
Since we need property 2 of definition 2 for our star product, we restrict ourselves
to graphs for which $B_{\Gamma}(1,\varphi) = B_{\Gamma}(\varphi,1) = 0$ i.e.
to graphs such that, for any terrestrial vertex $q_{j}$, at least one edge is
ending. Let us denote by $\tilde G_{n,m}$ the set of such graphs.
\begin{Definition} (K-star product)\\
A K-star product on $\R^{d}$ is a star product of the form:\\
$$\varphi \ast \psi = \varphi.\psi + \sum_{n\geq1} t^{n}\sum_{\Gamma\in 
\tilde G_{n,2}} a_{\Gamma} B_{\Gamma}(\Lambda,\Lambda,.....\Lambda)(\varphi\otimes\psi)$$\\
where $a_{\Gamma}$ is a constant.
\end{Definition}
{ \bf Remark:}
Kontsevich needs to eliminate the 
''small loops''
$\vec {p_{i}p_{i}}$ in order to define the form $\omega_{\Gamma}$, but he considers 
such a generalization for linear $\Lambda$ in \cite{r2}.\\

Up to the ordering of the aerial vertices, it is easy to consider all elements 
of $\tilde G_{2,2}$:\\
{\it {\bf Lemma:} (Description of $\tilde G_{2,2}$)\label{nad}\\
The set $\tilde G_{2,2}$ contains exactly 10 graphs} 

\begin{figure}[h]
\epsfxsize=15cm
$$
\epsfbox{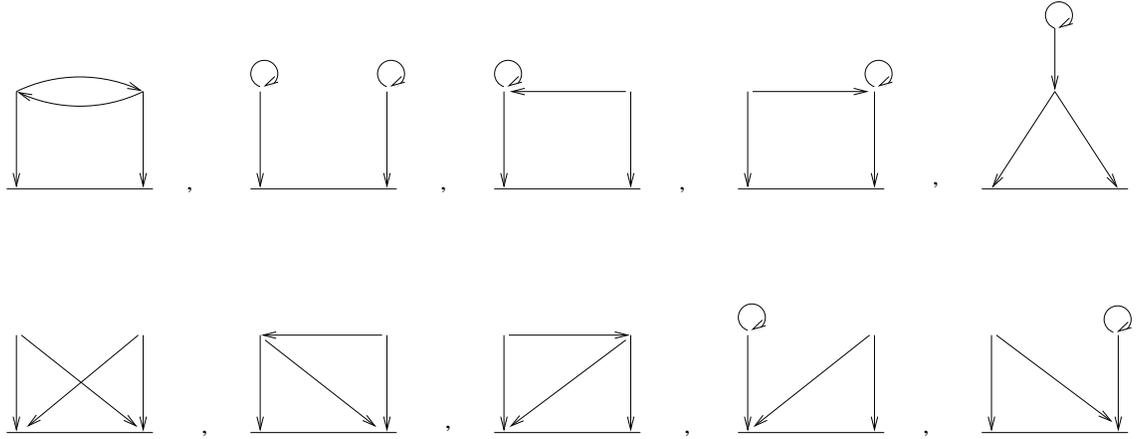}
$$
\caption{the $\tilde G_{2,2}$ elements}
\end{figure}

If we restrict ourselves to a symmetric $C_{2}$ in our star product, we have only to 
consider 6 graphs or linear combination of graphs 

\begin{figure}[htbp]
\epsfxsize=15cm
$$
\epsfbox{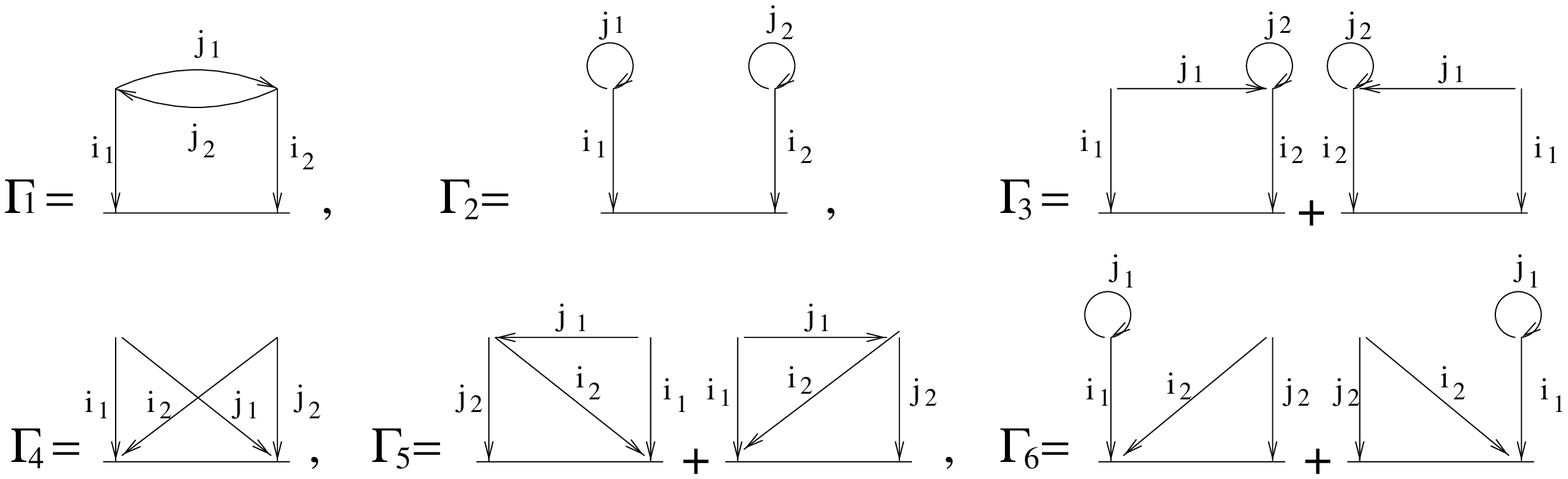}
$$
\caption{The symmetric elements of $\tilde G_{2,2}$}
\end{figure}

\pagebreak

\section{The quantum group $GL(2)$:}
Let us now consider the particular case of Lie group $G = GL(2) \subset \R^{4}$.
We endow $GL(2)$ with a Poisson-Lie structure by defining a $r$-matrix $\tilde r$ 
which verifies the modified CYBE:\\
$$ \tilde r = X_{+}\otimes X_{-} - X_{-}\otimes X_{+}~~~~ \in \wedge^{2}g $$\\
where $X_{+} = \left( \begin{array}{cc} 
                         0        &1  \\
                         0        &0  
                        \end{array}
                                    \right)$
and $X_{-} =\left( \begin{array}{cc} 
                         0        &0  \\
                         1        &0  
                        \end{array}
                                    \right)$.\\
So the corresponding Poisson bracket on $GL(2)$ has the following form:\\
$$ \{\varphi,\psi\} = X_{+}^{\ell}(\varphi)X_{-}^{\ell}(\psi) - X_{-}^{\ell}(\varphi)X_{+}^{\ell}(\psi) -
X_{+}^{r}(\varphi)X_{-}^{r}(\psi) + X_{-}^{r}(\varphi)X_{+}^{r}(\psi).$$\\
We consider the matrix $T = (t_{ij})_{i,j=1,2}$ of coordinate functions on $GL(2)$, i.e. the functions
$t_{ij}(g) = g_{ij}$, where, for $g\in G$, we denote by $g_{ij}$ its matrix elements.\\
Let us put:
$$ T=\left( \begin{array}{cc} 
                         t_{11}        &t_{12}  \\
                         t_{21}        &t_{22}  
                        \end{array}
                                    \right)
    =\left( \begin{array}{cc} 
                         a        &b  \\
                         c        &d  
                        \end{array}
                                    \right).$$\\
Left and Right actions of $G$ on matrix coordinates on $G$ are given by:

\begin{gather}
(X^{\ell}t_{ij})(g) = (gX)_{ij} = \sum_{k} t_{ik}(g)X_{kj} \notag \\
(X^{r}t_{ij})(g) = (Xg)_{ij} = \sum_{k} X_{ik}t_{kj}(g)
\end{gather}
with these notations, the Poisson bracket looks like:\\
$$ \Lambda^{ab} = \{a,b\} = ab ,~~\Lambda^{ac} = \{a,c\} = ac ,~~\Lambda^{bc} = \{b,c\} = 0$$ 
$$ \Lambda^{bd} = \{b,d\} = bd ,~~\Lambda^{cd} = \{c,d\} = cd ,~~\Lambda^{ad} = \{a,d\} = 2bc.$$\\
These relations define completely the Poisson-Lie group $GL(2)$ with $r$-matrix $\tilde r$ since
any $\varphi \in C^{\infty}(G)$ can be approximated by polynomial functions
in $a$, $b$, $c$, $d$.\\
Now, what about the quantization of this Poisson-Lie group i.e. how looks the star product in 
terms of coordinate functions?\\
In \cite{r1} L. Takhtajan gives an elegant form of his star product $(\ref{st})$:

\begin{equation}
T_{1}\ast T_{2} = F^{-1}T\otimes TF
\end{equation}
with $$T_{1} = T\otimes I$$
$$T_{2} = I\otimes T.$$\\
A solution of $(\ref{F})$ for $GL(2)$ was given by:
\begin{equation}\label{Fn}
F = e^{ \frac{-tP}{2}}\left( \begin{array}{cccc} 
                              \sqrt{q}  &0        &0        &0         \\
                              0         &u^{-1}   &0        &0         \\
                              0         &v        &u        &0         \\
                              0         &0        &0        &\sqrt{q}   
                               \end{array}
                           \right )                      
\end{equation}\\
where ~~$q = e^{t}, ~~~~u = \sqrt{\frac{2}{q+q^{-1}}},~~~~v = \frac{q-q^{-1}}
{\sqrt{2(q+q^{-1})}},$~~ 
and {\footnotesize $P$} is the permutation operator.\\
We shall call the corresponding star product the Takhtajan star product and 
denote it by $\ast_{T}$.
\begin{Proposition} (Computation of $\ast_{T}$){\rm \cite{r1}}\\
Taking the form $(\ref{Fn})$ of element F, we obtain the following relations:

\begin{equation}\label{e1}
\begin{array}{c}
\begin{split}
&a\ast_{T}a = a^{2},~~~~b\ast_{T}b = b^{2},~~~~c\ast_{T}c = c^{2},~~~~d\ast_{T}d = d^{2}\\
&a\ast_{T}b = \sqrt{\frac{2}{1+q^{-2}}}ab,~~~~a\ast_{T}c = \sqrt{\frac{2}{1+q^{-2}}}ac,~~~~b\ast_{T}c = \frac{2}{q+q^{-1}}bc\\
&b\ast_{T}d = \sqrt{\frac{2}{1+q^{-2}}}bd,~~~~c\ast_{T}d = \sqrt{\frac{2}{1+q^{-2}}}cd,~~~~a\ast_{T}d = ad + \frac{q-q^{-1}}{q+q^{-1}}bc.
\end{split}
\end{array}
\end{equation}
\end{Proposition}

\section{Comparing star product:}
We want now to compare the Takhtajan star product and the K-star product. We shall compare these
two sort of star products on a chart domain i.e.:
first we look at $GL(2)$ as an open subset of $\R^{4}$:\\
$$ GL(2) = \{T = \left( \begin{array}{cc}
                          a        &b  \\
                          c        &d   
                                 \end{array}
                                    \right),
~~~~ad-bc\neq0\} \subset \R^{4}=\{(a,b,c,d)\}$$\\
we call this chart the 
{\it''ordinary chart''}.
Then we look at the expontial mapping:

$$ exp: g\ell(2) = \{X = \left( \begin{array}{cc}
                             \alpha        &\beta  \\
                             \gamma        &\delta  \end{array}
                                    \right),~~~\Arrowvert X\Arrowvert < 2\pi \}
 \longrightarrow \{e^{X}\}\subset GL(2)$$
we call this chart the 
{\it''exponential chart''}.\\
The Takhtajan star product $\ast_{T}$ can be written on the ordinary or exponential chart as:

$$\varphi \ast_{T} \psi = \varphi.\psi + {t}C_{1}(\varphi,\psi) + t^{2}C_{T}(\varphi,\psi) 
+ \cdots $$\\
with $C_{T}$ symmetric.\\
Suppose that $\ast_{T}$ is a K-star product then $C_{T}$ has the form:
$$C_{T} = a_{\Gamma_{1}}B_{\Gamma_{1}}(\Lambda,\Lambda) + a_{\Gamma_{2}}B_{\Gamma_{2}}(\Lambda,\Lambda) 
+ \cdots + a_{\Gamma_{6}}B_{\Gamma_{6}}(\Lambda,\Lambda).$$
Computing $(\ref{e1})$, we find thus relations between the $a_{\Gamma_{i}}$ and it is 
possible to prove there is no solution for these relations.
We shall apply this method for the exponential chart.

Another possible way is to use the graph cohomology \cite{r4}. If we write the Kontsevich star product:
$$\varphi \ast_{K} \psi = \varphi.\psi + tC_{1}(\varphi,\psi) + t^{2}C_{K}(\varphi,\psi) + \cdots $$
Suppose that $\ast_{T}$ is a K-star product then $C_{K} - C_{T}$ being symmetric, 
is a coboundary $\delta T$, with $T = \sum_{n=1,2}t^{n}\sum_{\Gamma\in \tilde G_{n,1}}
K_{\Gamma}B_{\Gamma}(\Lambda, \Lambda)$.\\
We can compute $T$ and prove there is no solution again.
We shall apply this method for the ordinary chart.

\section{In the ordinary chart:}
On the Poisson-Lie group $GL(2) \subset \R^{4}$ we consider the chart 
$T =\left( \begin{array}{cc}
          a        &b  \\
          c        &d  \end{array}
                                    \right)$,
in this case we have:
$$\varphi \ast_{K} \psi = \varphi.\psi + tC_{1}(\varphi,\psi) + t^{2}C_{K}(\varphi,\psi) + \cdots$$
$$\varphi \ast_{T} \psi = \varphi.\psi + tC_{1}(\varphi,\psi) + t^{2}C_{T}(\varphi,\psi) + \cdots$$\\
where $C_{K}(C_{T})$ is the Kontsevich (the Takhtajan) bidifferential operator.\\
Since $ \Lambda $ is quadratic, if $ \varphi,\psi $ are coordinate functions, each term of these 
star products is quadratic.\\
Now $C_{K}$ and $C_{T}$ are symmetrics.\\
If we assume that we can write $C_{T}$ as:
 
$$C_{T} = \sum_{\Gamma\in \tilde G_{2,2}}a_{\Gamma}B_{\Gamma}(\Lambda,\Lambda)$$\\
then $C_{K}-C_{T}$ is a Hochschild cocycle which is symmetric and vanishing on 
constants, i.e. a coboundary, and there exists differential operators 
vanishing on constants $T_{1}$, $T_{2}$ such that:
\begin{equation}
T\varphi = \varphi + tT_{1}\varphi + t^{2}T_{2}\varphi
\end{equation}

satisfies:
\begin{equation}
T(\varphi \ast_{K} \psi) = T(\varphi) \ast_{T} T(\psi)
\end{equation}
and
\begin{figure}[htbp]
\epsfxsize=4cm
$$
\epsfbox{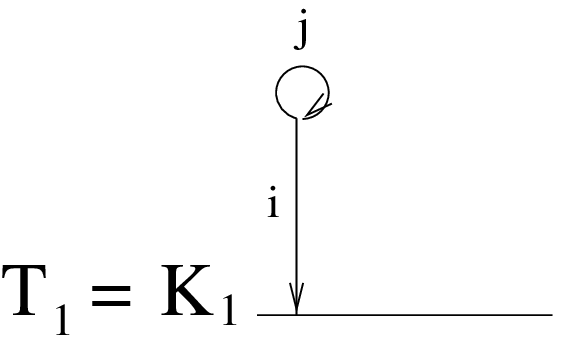}
$$
\end{figure}
\begin{figure}[htbp]
\epsfxsize=12cm
$$
\epsfbox{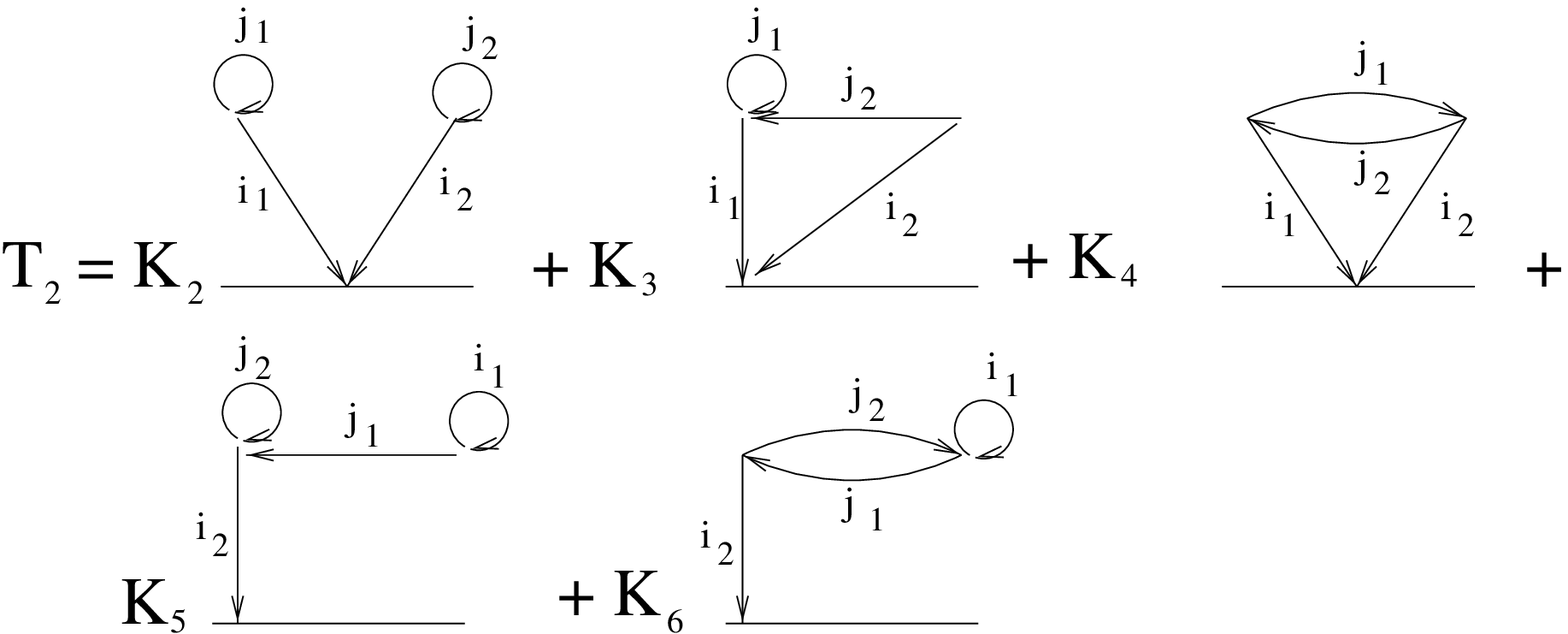}
$$
\end{figure}

then
\begin{equation}
\begin{array}{c}
\begin{split}
T &= Id + t\sum_{ij} K_{1}\partial_{j}\Lambda^{ij}\partial_{i} + 
t^{2}\sum_{i_{1}i_{2}j_{1}j_{2}} K_{2}
\partial_{j_{2}}\Lambda^{i_{2}j_{2}}\partial_{j_{1}}\Lambda^{i_{1}j_{1}}
\partial_{i_{1}}\partial_{i_{2}}\\
&+ t^{2}\sum_{i_{1}i_{2}j_{1}j_{2}} K_{3}
\Lambda^{i_{2}j_{2}}\partial_{j_{1}}\partial_{j_{2}}
\Lambda^{i_{1}j_{1}}\partial_{i_{1}}\partial_{i_{2}}
+ t^{2}\sum_{i_{1}i_{2}j_{1}j_{2}} K_{4}
\partial_{j_{1}}\Lambda^{i_{2}j_{2}}\partial_{j_{2}}\\
&\Lambda^{i_{1}j_{1}}\partial_{i_{1}}\partial_{i_{2}}
+ t^{2}\sum_{i_{1}i_{2}j_{1}j_{2}} K_{5}
\partial_{i_{1}}\Lambda^{i_{1}j_{1}}\partial_{j_{2}}\partial_{j_{1}}
\Lambda^{i_{2}j_{2}}\partial_{i_{2}}
+ t^{2}\sum_{i_{1}i_{2}j_{1}j_{2}} K_{6}\\
&\partial_{j_{2}}\partial_{j_{1}}\Lambda^{i_{1}j_{1}}\partial_{i_{1}}\Lambda^{i_{2}j_{2}}\partial_{i_{2}} 
\end{split}
\end{array}
\end{equation}\\
with the 4-upple of indexes $(i_{1},i_{2},j_{1},j_{2})\in \{a,b,c,d\}^{4}$.
So the equivalence between the two star products $\ast_{T}$ and $\ast_{K}$:
$$\sum_{p+q=2} T_{p}(C_{K})_{q}(\varphi,\psi) = \sum_{p+q+r=2} (C_{T})_{p}(T_{q}\varphi,T_{r}\psi)$$
gives the following system of equations:

\begin{equation}
\left \{ \begin{array}{c}
\begin{split}
     &2K_{3} + K_{4} = \frac{7}{48}\\
      &K_{3} + 2K_{4} = \frac{1}{6}\\
      &K_{4} - \frac{1}{12} = 0\\
      &K_{4} - \frac{1}{12}= -\frac{1}{8}\\
      &K_{1}^{2} + 2K_{2} + K_{4} = \frac{1}{12}.
\end{split}      
      \end{array}
      \right.
\end{equation}
This system has no solution.

\begin{Proposition} (Comparing on ordinary chart)\\
In the ordinary chart the Takhtajan star product can't be written as a K-star product.
\end{Proposition}

\section{In the exponential chart:}
Let us consider again the Poisson-Lie group $GL(2)$, but now, with an exponential chart 
$X = \left( \begin{array}{cc}
            \alpha        &\beta  \\
            \gamma        &\delta  \end{array}
                                    \right)$
such that:
$T =\left( \begin{array}{cc}
          a        &b  \\
          c        &d  \end{array}
                                    \right) = e^{X}$.\\
In this case we have:
\begin{equation}
\left \{ \begin{array}{c} 
\begin{split}
 &a = 1 + \alpha + \frac{\alpha^{2}+\beta\gamma}{2} + \frac{\alpha^{3}+\beta\gamma\alpha
+\beta\gamma\delta}{6} + \dots\\
&b = \beta + \frac{\beta \alpha +\beta \delta}{2} + \frac{\beta^{2} \gamma + \beta \alpha^{2}
+\beta\delta^{2}+\beta\alpha\delta}{6} + \dots\\
&c = \gamma + \frac{\gamma \alpha + \gamma \delta}{2} + \frac{\beta \gamma^{2} + \gamma \alpha^{2}
+\gamma \delta^{2}+ \gamma \alpha \delta}{6} + \dots\\
&d = 1 + \delta + \frac{\delta^{2}+ \beta \gamma}{2} + \frac{\delta^{3} + \beta \gamma \delta
+ \beta \gamma  \alpha}{6} + \dots 
\end{split}
\end{array}
\right. 
\end{equation}
and the Poisson structures up to third order looks as:

\begin{equation}
\left \{ \begin{array}{c} 
\begin{split}
   &\Lambda^{\alpha \beta} = \beta + \frac{1}{3} \beta^{2} \gamma + \frac{1}{3} 
   \beta \alpha^{2} + \dots\\
&\Lambda^{\alpha \gamma} = \gamma + \frac{1}{3}\beta\gamma^{2} + \frac{1}{3}
\gamma\alpha^{2} + \dots\\
&\Lambda^{\beta\delta} = \beta + \frac{1}{3}\beta^{2}\gamma + \frac{1}{3}\beta
\delta^{2} + \dots\\
&\Lambda^{\gamma\delta} = \gamma + \frac{1}{3}\beta\gamma^{2} + \frac{1}{3}
\gamma\delta^{2} + \dots\\
&\Lambda^{\beta\gamma} = 0\\
&\Lambda^{\alpha\delta} = \beta\gamma\alpha + \beta\gamma\delta + \dots  
\end{split}
\end{array}
 \right.
\end{equation}							 
If we try to write the Takhtajan star product as a K-star product we have to consider 
all symmetric graphs $\Gamma_{1},~\Gamma_{2},~\Gamma_{3},~\Gamma_{4},
~\Gamma_{5}$ and $\Gamma_{6}$, described in lemma (section~\ref{nad}).

We attribute respectly the weights $a_{\Gamma_{1}},~a_{\Gamma_{2}},~a_{\Gamma_{3}},~a_{\Gamma_{4}},
~a_{\Gamma_{5}}$ and $a_{\Gamma_{6}}$ to graphs $\Gamma_{1},~\Gamma_{2},~\Gamma_{3},~\Gamma_{4},
~\Gamma_{5}$ and $\Gamma_{6}$ such that the product:
$$ \varphi\ast\psi = \sum_{n=0} t^{n} \sum_{\Gamma\in\tilde G_{n,2}}a_{\Gamma}B_{\Gamma}(\Lambda,...,
\Lambda)(\varphi\otimes\psi)$$
is associative.\\
So we calculate the operator:

\begin{equation}
\begin{array}{c}
\begin{split}
&\sum_{i_{1}i_{2}j_{1}j_{2}} a_{\Gamma_{1}}
\partial_{j_{1}}\Lambda^{i_{2}j_{2}}\partial_{j_{2}}\Lambda^{i_{1}j_{1}}\partial_{i_{1}}\otimes
\partial_{i_{2}}
+ \sum_{i_{1}i_{2}j_{1}j_{2}}a_{\Gamma_{2}}
\partial_{j_{1}}\Lambda^{i_{1}j_{1}}\partial_{j_{2}}\Lambda^{i_{2}j_{2}}\partial_{i_{1}}\otimes
\partial_{i_{2}}\\
&+ \sum_{i_{1}i_{2}j_{1}j_{2}}a_{\Gamma_{3}}
\Lambda^{i_{1}j_{1}}
\partial_{j_{1}}\partial_{j_{2}}\Lambda^{i_{2}j_{2}}(\partial_{i_{1}}
\otimes\partial_{i_{2}}+\partial_{i_{2}}\otimes\partial_{i_{1}})
+ \sum_{i_{1}i_{2}j_{1}j_{2}}a_{\Gamma_{4}}
\Lambda^{i_{1}j_{1}}\Lambda^{i_{2}j_{2}}\\
&\partial_{i_{2}}\partial_{i_{1}}\otimes\partial_{j_{2}}\partial_{j_{1}} 
+ \sum_{i_{1}i_{2}j_{1}j_{2}}a_{\Gamma_{5}}
\Lambda^{i_{1}j_{1}}\partial_{j_{1}}\Lambda^{i_{2}j_{2}}(\partial_{i_{2}}
\partial_{i_{1}}\otimes\partial_{j_{2}}+\partial_{j_{2}}\otimes\partial_{i_{2}}\partial_{i_{1}})\\
&+ \sum_{i_{1}i_{2}j_{1}j_{2}}a_{\Gamma_{6}}
\Lambda^{i_{2}j_{2}}\partial_{j_{1}}\Lambda^{i_{1}j_{1}}(\partial_{i_{2}}
\partial_{i_{1}}\otimes\partial_{j_{2}}+\partial_{j_{2}}\otimes\partial_{i_{2}}\partial_{i_{1}})
\end{split}
\end{array}
\end{equation}\\
associated to the graphs of (fig. 2),
on each pair $(\varphi,\psi)$ of functions $\varphi, \psi \in \{a,b,c,d\}$ and 4-upple of indexes
$(i_{1},i_{2},j_{1},j_{2}) \in \{\alpha,\beta,\delta,\gamma\}^{4}$.\\
Then the vanishing of the bidifferential operator $C_{K}-C_{T}$ gives this 
system of equations:

\begin{equation}
\left \{ \begin{array}{c}
\begin{split} 
         &a_{\Gamma_{1}} + 2a_{\Gamma_{2}}  = 0 \\
        &10a_{\Gamma_{1}} + 32a_{\Gamma_{2}} + 28a_{\Gamma_{3}} - 
         6a_{\Gamma_{4}} - 8a_{\Gamma_{5}} - 16a_{\Gamma_{6}}  = 0 \\
       &8a_{\Gamma_{3}} + a_{\Gamma_{5}} + 4a_{\Gamma_{6}} = 0 \\
       &8a_{\Gamma_{3}} + 2a_{\Gamma_{5}} + 4a_{\Gamma_{6}} = -\frac{3}{2} \\
       &a_{\Gamma_{5}} + 2a_{\Gamma_{6}}  = -\frac{1}{8} \\
       &-a_{\Gamma_{1}} - 2a_{\Gamma_{2}} - 6a_{\Gamma_{3}} + 
         4a_{\Gamma_{5}} + 8a_{\Gamma_{6}}  = -\frac{9}{16} \\
       &-a_{\Gamma_{1}} + 2a_{\Gamma_{2}} + 2a_{\Gamma_{3}} + 
         2a_{\Gamma_{5}} + 4a_{\Gamma_{6}}  = -\frac{3}{16}
\end{split} 
\end{array}\right.
\end{equation}\\
wich is a system with no solution.

For instance let us give  the calculation, up to second order, of $C_{K}(a,d)$ as an example from
which we obtain the first and second equations.\\
We determine $B_{\Gamma_{n}}(a,d)$ for $n=1,2,...,6$. We get for 

\begin{figure}[htbp]
\epsfxsize=3cm
$$
\epsfbox{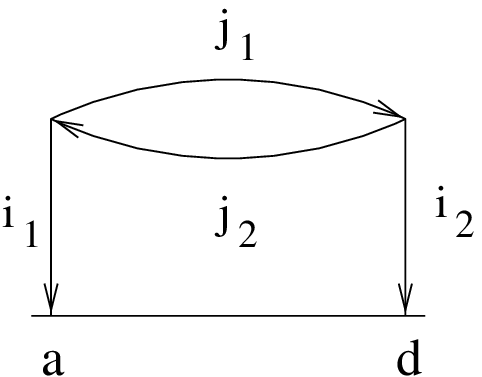}
$$
\caption{$\Gamma_{1}$}
\end{figure}

the functions:\\
$$B_{\Gamma_{1}}(a,d) = \sum_{i_{1}i_{2}j_{1}j_{2}}
\partial_{j_{1}}\Lambda^{i_{2}j_{2}}\partial_{j_{2}}\Lambda^{i_{1}j_{1}}\partial_{i_{1}}a
 \partial_{i_{2}}d $$\\
\noindent{\bf case1:}

$\left \{ \begin{array}{c} 
         i_{1}=\alpha \\
        j_{1}=\beta 
	  \end{array}
	  \right.$\\
So if we calculate ~~$\partial_{\beta}\Lambda^{i_{2}j_{2}}
\partial_{j_{2}}\Lambda^{\alpha\beta}\partial_{\alpha}a\partial_{i_{2}}d $,
with $i_{2},j_{2} \in \{\alpha,\beta,\gamma,\delta\}$, we get:
$$(\partial_{\beta}\Lambda^{i_{2}j_{2}}
\partial_{j_{2}}\Lambda^{\alpha\beta}\partial_{\alpha}a
\partial_{i_{2}}d)_{i_{2}j_{2} = \alpha,\beta,\gamma,\delta} = 
-1 - \alpha - \delta - \alpha\delta -\frac{5}{6}\alpha^{2} - 
\frac{5}{6}\delta^{2} - \frac{11}{6}\beta\gamma + 0(2)$$\\ 
\noindent{\bf case2:}

$\left \{ \begin{array}{c}
        i_{1}=\beta\\
        j_{1}=\alpha\end{array} \right.~~~
(\partial_{\alpha}\Lambda^{i_{2}j_{2}}
\partial_{j_{2}}\Lambda^{\beta\alpha}\partial_{\beta}a\partial_{i_{2}}d) 
_{i_{2}j_{2} = \alpha,\beta,\gamma,\delta} = 0(2)$\\

\noindent{\bf case3:}

$\left \{ \begin{array}{c}i_{1}=\alpha\\
        j_{1}=\gamma\end{array} \right.~~~
(\partial_{\gamma}\Lambda^{i_{2}j_{2}}
\partial_{j_{2}}\Lambda^{\alpha\gamma}\partial_{\alpha}a\partial_{i_{2}}
d )_{i_{2}j_{2} = \alpha,\beta,\gamma,\delta} = 
-1 - \alpha - \delta - \alpha\delta -\frac{5}{6}\alpha^{2} - 
\frac{5}{6}\delta^{2} - \frac{11}{6}\beta\gamma + 0(2)$\\

\noindent{\bf case4:}

$\left \{ \begin{array}{c}i_{1}=\gamma\\
        j_{1}=\alpha\end{array} \right.~~~
(\partial_{\alpha}\Lambda^{i_{2}j_{2}}
\partial_{j_{2}}\Lambda^{\gamma\alpha}\partial_{\gamma}a\partial_{i_{2}}d) 
_{i_{2}j_{2} = \alpha,\beta,\gamma,\delta} = 0(2)$\\

\noindent{\bf case5:}

$\left \{ \begin{array}{c}i_{1}=\beta\\
        j_{1}=\delta\end{array} \right.~~~
(\partial_{\delta}\Lambda^{i_{2}j_{2}}
\partial_{j_{2}}\Lambda^{\beta\delta}\partial_{\beta}a\partial_{i_{2}}d) 
_{i_{2}j_{2} = \alpha,\beta,\gamma,\delta} = 0(2)$\\

\noindent{\bf case6:}

$\left \{ \begin{array}{c}i_{1}=\delta\\
        j_{1}=\beta\end{array} \right.~~~
(\partial_{\beta}\Lambda^{i_{2}j_{2}}
\partial_{j_{2}}\Lambda^{\delta\beta}\partial_{\delta}a\partial_{i_{2}}d) 
_{i_{2}j_{2} = \alpha,\beta,\gamma,\delta} = \frac{1}{6}\beta\gamma + 0(2)$\\

\noindent{\bf case7:}

$\left \{ \begin{array}{c}i_{1}=\gamma\\
        j_{1}=\delta\end{array} \right.~~~
(\partial_{\delta}\Lambda^{i_{2}j_{2}}
\partial_{j_{2}}\Lambda^{\gamma\delta}\partial_{\gamma}a\partial_{i_{2}}d) 
_{i_{2}j_{2} = \alpha,\beta,\gamma,\delta} = 0(2)$\\

\noindent{\bf case8:}

$\left \{ \begin{array}{c}i_{1}=\delta\\
        j_{1}=\gamma\end{array} \right.~~~
(\partial_{\gamma}\Lambda^{i_{2}j_{2}}
\partial_{j_{2}}\Lambda^{\delta\gamma}\partial_{\delta}a\partial_{i_{2}}d) 
_{i_{2}j_{2} = \alpha,\beta,\gamma,\delta} = \frac{1}{6}\beta\gamma + 0(2)$\\

\noindent{\bf case9:}

$\left \{ \begin{array}{c}i_{1}=\alpha\\
        j_{1}=\delta\end{array} \right.~~~
(\partial_{\delta}\Lambda^{i_{2}j_{2}}
\partial_{j_{2}}\Lambda^{\alpha\delta}\partial_{\alpha}a\partial_{i_{2}}d) 
_{i_{2}j_{2} = \alpha,\beta,\gamma,\delta} = 0(2)$\\

\noindent{\bf case10:}

$\left \{ \begin{array}{c}i_{1}=\delta\\
        j_{1}=\alpha\end{array} \right.~~~
(\partial_{\alpha}\Lambda^{i_{2}j_{2}}
\partial_{j_{2}}\Lambda^{\delta\alpha}\partial_{\delta}a\partial_{i_{2}}d) 
_{i_{2}j_{2} = \alpha,\beta,\gamma,\delta} = 0(2)$\\

then we have:
$$B_{\Gamma_{1}}(a,d) = - 2 - 2\alpha - 2\delta - 2\alpha\delta - \frac{5}{3}\alpha^{2}
- \frac{5}{3}\delta^{2} - \frac{10}{3}\beta\gamma + 0(2).$$\\
Similarly we calculate :
\begin{figure}[htbp]
\epsfxsize=3cm
$$
\epsfbox{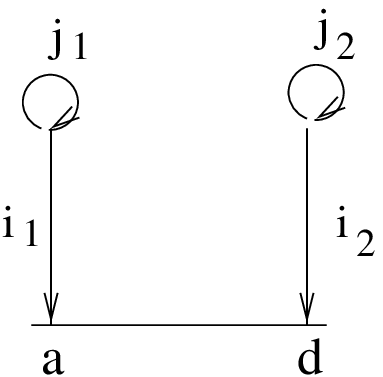}
$$
\caption{$\Gamma_{2}$}
\end{figure}

$$B_{\Gamma_{2}}(a,d) = \sum_{i_{1}i_{2}j_{1}j_{2}}
\partial_{j_{2}}\Lambda^{i_{2}j_{2}}\partial_{j_{1}}\Lambda^{i_{1}j_{1}}\partial_{i_{1}}a
 \partial_{i_{2}}d $$\\

thus we have:
$$B_{\Gamma_{2}}(a,d) =  - 4 - 4\alpha - 4\delta - 4\alpha\delta - \frac{10}{3}\alpha^{2}
- \frac{10}{3}\delta^{2} - \frac{32}{3}\beta\gamma + 0(2)$$\\
and for:

\begin{figure}[htbp]
\epsfxsize=7cm
$$
\epsfbox{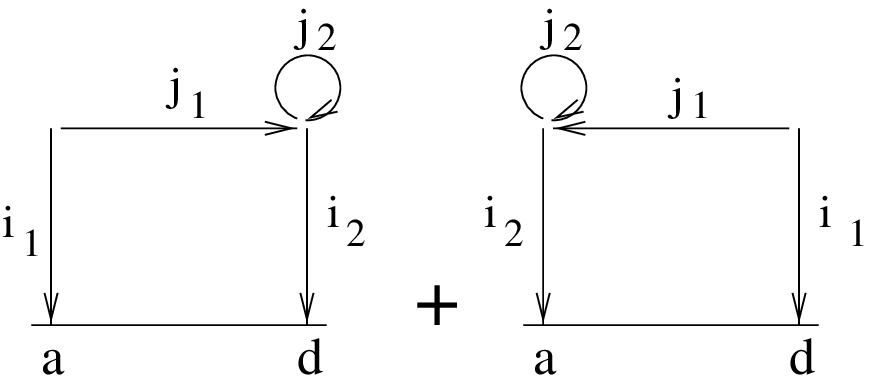}
$$
\caption{$\Gamma_{3}$}
\end{figure}

$$B_{\Gamma_{3}}(a,d) = \sum_{i_{1}i_{2}j_{1}j_{2}}
\partial_{j_{2}}\partial_{j_{1}}\Lambda^{i_{2}j_{2}}\Lambda^{i_{1}j_{1}}
(\partial_{i_{1}}a\partial_{i_{2}}d + \partial_{i_{2}}a\partial_{i_{1}}d)$$\\

then we have:
$$B_{\Gamma_{3}}(a,d) = - \frac{28}{3}\beta\gamma + 0(2)$$\\
and:

\begin{figure}[htbp]
\epsfxsize=3cm
$$
\epsfbox{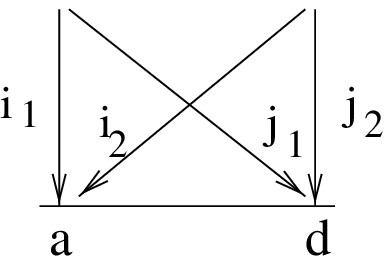}
$$
\caption{$\Gamma_{4}$}
\end{figure}

$$B_{\Gamma_{4}}(a,d) = \sum_{i_{1}i_{2}j_{1}j_{2}}
\Lambda^{i_{2}j_{2}}\Lambda^{i_{1}j_{1}}
\partial_{i_{2}}\partial_{i_{1}}a\partial_{j_{2}}\partial_{j_{1}}d$$\\
$$B_{\Gamma_{4}}(a,d) = 2\beta\gamma + 0(2)$$\\
and:

\begin{figure}[htbp]
\epsfxsize=7cm
$$
\epsfbox{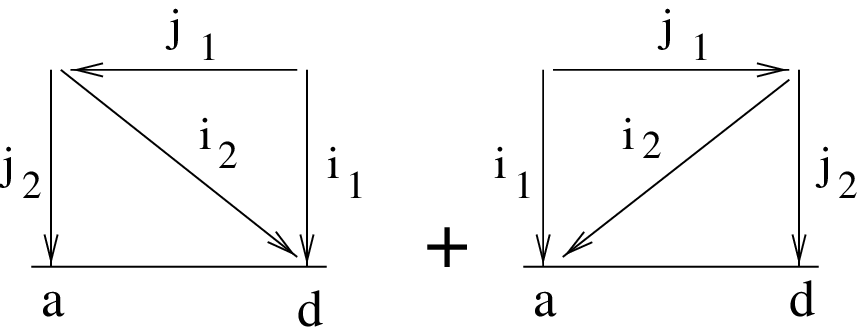}
$$
\caption{$\Gamma_{5}$}
\end{figure}
$$B_{\Gamma_{5}}(a,d) = \sum_{i_{1}i_{2}j_{1}j_{2}}
\partial_{j_{1}}\Lambda^{i_{2}j_{2}}\Lambda^{i_{1}j_{1}}
(\partial_{j_{2}}a\partial_{i_{2}}\partial_{i_{1}}d + \partial_{i_{2}}
\partial_{i_{1}}a\partial_{j_{2}}d)$$\\
$$B_{\Gamma_{5}}(a,d) = \frac{8}{3}\beta\gamma + 0(2)$$\\
and:

\begin{figure}[htbp]
\epsfxsize=7cm
$$
\epsfbox{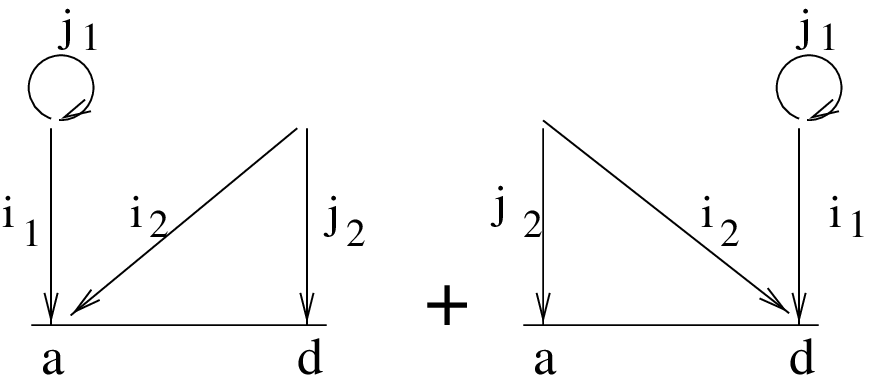}
$$
\caption{$\Gamma_{6}$}
\end{figure}
$$B_{\Gamma_{6}}(a,d) = \sum_{i_{1}i_{2}j_{1}j_{2}}
\Lambda^{i_{2}j_{2}}\partial_{j_{1}}\Lambda^{i_{1}j_{1}}
(\partial_{i_{2}}\partial_{i_{1}}a\partial_{j_{2}}d + \partial_{j_{2}}a
\partial_{i_{2}}\partial_{i_{1}}d)$$\\
$$B_{\Gamma_{6}}(a,d) = \frac{16}{3}\beta\gamma + 0(2).$$\\
Now considering weights $a_{\Gamma_{1}}, a_{\Gamma_{2}}, a_{\Gamma_{3}}, a_{\Gamma_{4}}, a_{\Gamma_{5}}$
and $a_{\Gamma_{6}}$, we have:\\

\begin{equation*}
\begin{split}
C_{K}(a,d) &= (-2a_{\Gamma_{1}} - 4a_{\Gamma_{2}}) + (-2a_{\Gamma_{1}}- 4a_{\Gamma_{2}})
\alpha + (-2a_{\Gamma_{1}}- 4a_{\Gamma_{2}})\delta\\ 
&+ (-2a_{\Gamma_{1}}- 4a_{\Gamma_{2}})
\alpha\delta + (-\frac{5}{3}a_{\Gamma_{1}} - \frac{10}{3}a_{\Gamma_{2}})\alpha^{2}
+ (-\frac{5}{3}a_{\Gamma_{1}} - \frac{10}{3}a_{\Gamma_{2}})\delta^{2} +\\
&(-\frac{10}{3}a_{\Gamma_{1}} - \frac{32}{3}a_{\Gamma_{2}} - \frac{28}{3}a_{\Gamma_{3}} + 
2a_{\Gamma_{4}} + \frac{8}{3}a_{\Gamma_{5}} + \frac{16}{3}a_{\Gamma_{6}})\beta\gamma.
\end{split}
\end{equation*}\\
In the other hand, we have:\\
$$C_{T}(a,d) = 0$$\\
then we obtain our two first equations.\\
In the same way we obtain the remaining equations.

\begin{Proposition} (Comparing on exponential chart)\\
In the exponential chart we can't write the Takhtajan star product as a K-star product.

\end{Proposition}
{\bf Acknowledgements}\\
This paper owes its existence to D. Arnal, I am very grateful to him for help
and guidance.

\end{document}